\documentclass[11pt,epsf]{article}

\usepackage{epsfig}
\usepackage{amssymb}
\usepackage{graphicx}
\usepackage{color}

\begin{document}

\baselineskip 6mm
\renewcommand{\thefootnote}{\fnsymbol{footnote}}


\newcommand{\nc}{\newcommand}
\newcommand{\rnc}{\renewcommand}



\newcommand{\tcb}{\textcolor{blue}}
\newcommand{\tcr}{\textcolor{red}}
\newcommand{\tcg}{\textcolor{green}}


\def\be{\begin{equation}}
\def\ee{\end{equation}}
\def\ba{\begin{array}}
\def\ea{\end{array}}
\def\bea{\begin{eqnarray}}
\def\eea{\end{eqnarray}}
\def\nn{\nonumber\\}


\def\ct{\cite}
\def\la{\label}
\def\eq#1{(\ref{#1})}


\def\a{\alpha}
\def\b{\beta}
\def\g{\gamma}
\def\G{\Gamma}
\def\d{\delta}
\def\D{\Delta}
\def\e{\epsilon}
\def\et{\eta}
\def\ph{\phi}
\def\Ph{\Phi}
\def\ps{\psi}
\def\Ps{\Psi}
\def\k{\kappa}
\def\l{\lambda}
\def\L{\Lambda}
\def\m{\mu}
\def\n{\nu}
\def\th{\theta}
\def\Th{\Theta}
\def\r{\rho}
\def\s{\sigma}
\def\S{\Sigma}
\def\ta{\tau}
\def\o{\omega}
\def\O{\Omega}
\def\pr{\prime}


\def\half{\frac{1}{2}}

\def\goto{\rightarrow}

\def\na{\nabla}
\def\grad{\nabla}
\def\curl{\nabla\times}
\def\div{\nabla\cdot}
\def\pa{\partial}

\def\bra{\left\langle}
\def\ket{\right\rangle}
\def\lb{\left[}
\def\lc{\left\{}
\def\ls{\left(}
\def\lp{\left.}
\def\rp{\right.}
\def\rb{\right]}
\def\rc{\right\}}
\def\rs{\right)}
\def\fr{\frac}
\def\text#1{{\rm #1}}

\def\vac#1{\mid #1 \rangle}


\def\text#1{{\rm #1}}
\def\td#1{\tilde{#1}}
\def\check{ \maltese {\bf Check!}}


\def\Tr{{\rm Tr}\,}
\def\det{{\rm det}}


\def\bc#1{\nnindent {\bf $\bullet$ #1} \\ }
\def\ch {$<Check!>$ }
\def\ss {\vspace{1.5cm}}

\begin{titlepage}

\hfill\parbox{5cm} {APCTP Pre2016-009}

\vspace{25mm}

\begin{center}
{\Large \bf Thermodynamic law from the entanglement entropy bound}

\vskip 1. cm
  {Chanyong Park$^{a,b}$\footnote{e-mail : chanyong.park@apctp.org}}

\vskip 0.5cm

{\it $^a\,$ Asia Pacific Center for Theoretical Physics, Pohang, 790-784, Korea } \\
{\it $^b\,$ Department of Physics, Postech, Pohang, 790-784, Korea }\\

\end{center}

\thispagestyle{empty}

\vskip2cm


\centerline{\bf ABSTRACT} \vskip 4mm

\vspace{1cm}

From black hole thermodynamics, the Bekenstein bound has been proposed as a universal thermal entropy bound. It has been further generalized to an entanglement entropy bound which is valid even in a quantum system. In a quantumly entangled system, the non-negativity of the relative entropy leads to the entanglement entropy bound. When the entanglement entropy bound is saturated, a quantum system satisfies the thermodynamics-like law with an appropriately defined entanglement temperature. We show that the saturation of the entanglement entropy bound accounts for a universal feature of the entanglement temperature proportional to the inverse of the system size. In addition, we show that the deformed modular Hamiltonian under a global quench also satisfies the generalized entanglement entropy boundary after introducing a new quantity called the entanglement chemical potential.

\vspace{2cm}


\end{titlepage}

\renewcommand{\thefootnote}{\arabic{footnote}}
\setcounter{footnote}{0}



\section{Introduction}

Recently, much attention has been paid  to the entanglement entropy for studying various quantum properties in string theory as well as condensed matter physics. To describe the entanglement entropy of a strongly interacting system, it has been proposed  that its entanglement entropy can be evaluated by calculating the minimal surface area in the dual gravity theory according to the AdS/CFT correspondence \cite{Ryu:2006bv,Ryu:2006ef}. For a two-dimensional conformal field theory (CFT) whose dual gravity is represented as a three-dimensional AdS geometry, the holographic entanglement entropy calculation has exactly reproduced the two-dimensional CFT results  \cite{Calabrese:2004eu,Calabrese:2005zw,Calabrese:2009qy}. This work has been further generalized to higher dimensional theories and non-conformal cases \cite{Solodukhin:2006xv}-\cite{Park:2015dia}. Intriguingly, it has been shown that the entanglement entropy of excited systems satisfies the first law of thermodynamics after defining an entanglement temperature appropriately. Moreover, the entanglement temperature shows a universal behavior proportional to the inverse of the system size \cite{Bhattacharya:2012mi,Bianchi:2012ev,Nozaki:2013vta,Allahbakhshi:2013rda,Wong:2013gua,Momeni:2015vka}. This universality has also been checked in various hyperscaling violation geometries \cite{He:2013rsa,Pang:2013lpa,Park:2015afa}.

In black hole physics, the Bekenstein bound has been proposed from a thought experiment \cite{Bekenstein:1973ur,Bekenstein:1974ax,Bekenstein:1980jp,Unruh:1982ic,Unruh:1983ir}. When an object is absorbed into a black hole, it describes that the increased thermal entropy of a black hole should be bounded from the absorbed energy. This Bekenstein bound can be further generalized to an entanglement entropy bound which is also valid in quantum systems. When a ground state is excited, the entanglement entropy bound implies that the increased entanglement entropy is bounded by the excitation energy similar to the Bekenstein bound. This entanglement entropy bound has been derived from the non-negativity of the relative entropy \cite{Casini:2008cr,Blanco:2013joa}. The non-negativity of the relative entropy comes from the fact that the vacuum or thermal state has a minimum entanglement or thermal entropy respectively. In general, the entanglement entropy bound is saturated only when two states are equal. In the UV limit, however,  the increased entanglement entropy with only lower order corrections can saturate the entanglement entropy bound. This is the dominant contribution associated with the thermodynamics-like law. Then the non-negativity of the relative entropy implies that the ignored small higher order corrections should be negative.

In this work, we will study the entanglement entropy of a quantum system with excitations and/or a global quench and then investigate the universality of its thermodynamics-like law. To do so, we take into account a charged AdS black brane geometry with peculiar properties \cite{Gubser:2009qt,Gubser:2000mm,Davison:2013uha}. This geometry allows for the dual field theory to have a Fermi sea and massless fluctuations on the Fermi surface.  Unlike the RNAdS black brane having a nonzero Bekenstein-Hawking entropy even at zero temperature, it has zero Bekenstein-Hawking entropy at zero temperature. Using the holographic renormalization technique, its thermodynamic properties have been studied from the boundary stress tensor \cite{Park:2014gja}. Interestingly, it has been shown that the trace of the boundary stress tensor does not vanish even though it has an AdS asymptote. This fact implies that matter of the dual field theory is non-conformal, while matter in the dual QFT of the RNAdS black brane is conformal. The geometry we consider provides an interesting background to study a strongly interacting fermionic system. On this interesting background, we investigate the entanglement entropy bound. We explicitly show that the lower order entanglement entropies in strip- and ball-shaped regions saturate the entanglement entropy bound as mentioned before. 

When a thermal system is deformed by a chemical potential or number of particles, its thermodynamic law is generally modified due to an additional conserved quantity. This also happens in the entanglement entropy bound \cite{Park:2015afa}. Assume that a CFT theory is deformed by a certain global quench. If a global quench is relevant, its effect becomes small at least in the UV limit. In this case, we can study this system by using the perturbation of the CFT. In general, a global quench modifies the modular Hamiltonian whose relative entropy leads to a generalized entanglement entropy. In this case like the thermal system, one  can introduce a new parameter called the entanglement chemical potential in order to describe the entanglement entropy change caused by the global quench. We show that the generalized entanglement entropy bound after a global quench still satisfies the generalized thermodynamics-like law.

The rest of this paper is organized as follows. In section 2, we discuss the general aspects of the entanglement entropy bound and the universality of the entanglement temperature. After reviewing thermodynamic properties of a charged black brane in section 3, we explicitly show in section 4 that the lower order entanglement entropies of an excited state  saturate the entanglement entropy bound, and that their thermodynamic interpretation leads to a universal entanglement temperature. We also discuss the effect of a global quench which satisfies the generalized entanglement entropy bound. We finish our work with some concluding remarks in section 5.


\section{Entanglement entropy bound}

The Bekenstein bound has been proposed as a universal bound of the thermal entropy in flat space. It has been originally conceived through a thought experiment for black hole thermodynamics and classical physics \cite{Bekenstein:1973ur,Bekenstein:1974ax,Bekenstein:1980jp}. When an object is absorbed into a black hole, the entropy of an object increases the black hole area due to the generalized second law of thermodynamics. This is in turn governed by the Einstein equations and implies that the increased entropy is bounded by the absorbed energy
\be
\D S \le \l  l \D E,
\ee
where $l$ and $\l$ are a typical size of the system and a non-universal numerical factor of order $1$. The Bekenstein bound is universal in that it is independent of microscopic details up to $\l$. Recently, the entanglement entropy has been proposed as the origin of black hole entropy \cite{Bombelli:1986rw,Srednicki:1993im,Frolov:1993ym,Susskind:1994sm,Solodukhin:2011gn}. In the entanglement entropy context, it has been argued that a generalized Bekenstein bound can be also applied to a quantum system \cite{Casini:2008cr,Blanco:2013joa}.

To understand such a generalized Bekenstein bound, we need to define a relative entropy which is independent of the renormalization scheme. When two states are in the same Hilbert space, the relative entropy gives rise to a fundamental statistical measure of their distance. If two reduced density matrices are denoted by $\r_1$ and $\r_0$, the relative entropy $S(\r_1 | \r_0)$ is defined as 
\be
S(\r_1 | \r_0) \equiv \Tr (\r_1 \log \r_1) - \Tr (\r_1 \log \r_0) .
\ee
Here we can identify $\r_0$ with the reduced density matrix of a ground or thermal state, while $\r_1$ is one for a quantumly or thermally excited state. If there exists a parameter connecting two reduced density matrices such that $\r_1=\r_1 (\l)$ and  $\r_0 = \r_1 (0)$, the relative entropy usually has a non-negativity value
\be			
S(\r_0 | \r_0) = 0 \quad {\rm and} \quad S(\r_1 | \r_0) > 0 \ {\rm for} \ \r_0 \ne \r_1 .
\ee
Thus, $\r_0$ corresponds to a minimum point \cite{Blanco:2013joa}. Using the definition of the entanglement entropy, the relative entropy can be reexpressed as
\be
S(\r_1 | \r_0) = \D  K  - \D S ,
\ee
where variations of the modular Hamiltonian and entanglement entropy are given by
\be
\D   K   = \Tr (\r_1 K) - \Tr (\r_0 K) \quad {\rm and} \quad \D S = S (\r_1) - S (\r_0) .
\ee
The non-negativity of the relative entropy leads to the following relation
\be
\D   K   \ge \D S ,
\ee
which has been regarded as a generalized Bekenstein bound holding for any region in QFT. From now on, we call it an entanglement entropy bound. As will be demonstrated, the entanglement entropy bound is equivalent to the Bekenstein bound except that it is also working in a quantum system. Note that the exact saturation of the entanglement entropy bound occurs only when $\r_0 = \r_1$. In a UV limit, however, we can define an {\it almost} saturated entanglement entropy bound, $\D \bra K \ket \approx \D S$. To clarify the meaning of `{\it almost}', we first note that the increased entanglement entropy in a UV limit can be divided into two parts, a dominant part and higher order corrections,
\be
\D S = \D S_{dom} +  \D S_{high} .
\ee
Ignoring higher order corrections, the dominant part can satisfy $\D   K   = \D S_{dom}$. Therefore, the almost saturated entanglement entropy bound implies that the entanglement entropy bound is saturated up to higher order corrections. For consistency, higher order corrections should be negative, $\D S_{high}<0$. The almost saturated entanglement entropy bound is important to understand universal features of an entangled quantum system and its thermodynamics-like law. In addition, it has been used to reconstruct the linearized Einstein equation for the AdS geometry \cite{Swingle:2014uza,Lashkari:2013koa,Faulkner:2013ica}.

In recent studies \cite{Bhattacharya:2012mi,Bianchi:2012ev,Nozaki:2013vta}, it has been found that the entanglement entropy of an excited state in a strip region follows the thermodynamics-like law after defining an appropriate entanglement temperature. Intriguingly, the entanglement temperature shows a universal feature inversely proportional to the strip width. Now we can ask whether the similar universal feature also occurs in the entanglement entropy involved in a different shaped region and why such a universal feature occurs. The goal of this work is to answer these questions by using the entanglement entropy bound. Before describing the details, we give a general argument on the universality of the entanglement temperature. Following the simple dimension counting in a relativistic QFT, we can guess that the increased modular Hamiltonian is proportional to the increased energy
\be		\la{ass:formofmodHam}
\D   K   = \l \ l \D E
\ee
with a non-universal numerical factor $\l$. In general, the modular Hamiltonian is not known except several simple cases. One of them is the case with a spherical entangling surface. In section 4, we will show that the modular Hamiltonian in a ball-shaped region really satisfies \eq{ass:formofmodHam}. Substituting \eq{ass:formofmodHam} into the entanglement entropy bound, we finally arrive at the Bekenstein bound working in a QFT. When the entanglement entropy bound is almost saturated, $\D K \approx \D S$, we can reinterpret it as the thermodynamics-like law. In this case, the entanglement temperature has the following form,
\be
T_E \equiv \fr{\D E}{\D S} = \fr{1}{\l  l} ,
\ee
where $\l$ depends on the shape of the entangling surface but not the system size. Focusing on the size dependence, the entanglement temperature is proportional to the inverse of the system size, $T_E \sim 1/l$, regardless of the shape of the entangling surface and details of the underlying theory \cite{Bhattacharya:2012mi,Blanco:2013joa}. This feature is similar to the universality of the Bekenstein bound, so we can call it the universality of the entanglement temperature. In the next sections, we will study the entanglement entropy of a holographic fermion system dual to an Einstein-Maxwell-scalar gravity and explicitly show that the almost entanglement entropy bound leads to the universal entanglement temperature. 

\section{Charged black brane with a scalar hair}

Recently, numerous charged dilatonic black brane solutions have been extensively studied for understanding the Fermi surface \cite{Gouteraux:2011ce,Dong:2012se,Kulkarni:2012re,Kulkarni:2012in,Park:2013ana,He:2013rsa,Pang:2013lpa,Park:2015afa}. In \cite{Gubser:2009qt}, it has been shown that  a Fermi surface and massless fluctuations on it can be described by the following dual gravity,
\be
S = \fr{1}{2 \k^2} \int d^5 x \sqrt{-g} \lb {\cal R} - \fr{1}{4} e^{4 \ph} F_{\m\n} F^{\m\n}
- 12 \pa_{\m} \ph  \pa^{\m} \ph + \fr{1}{R^2} \ls 8 e^{2 \ph} + 4 e^{- 4 \ph} \rs
\rb ,
\ee
where $R=1$ corresponds to an AdS radius. This action shows that there is a local minimum at $\ph=0$ where the scalar potential reproduces the five-dimensional AdS cosmological constant. If there exists a nontrivial $\ph$ approaching to zero at the asymptotic boundary, there exists an asymptotic AdS solution. Solving equations of motion gives rise to the following solution
\bea
ds^2 &=& r^2  e^{2 A(r)}  \ls - f(r) dt^2 + d \vec{x}^2 \rs + \fr{e^{2 B(r)} }{r^2 f(r)} dr^2 , \nn
A &=& A_t dt ,
\eea
with
\bea			\la{sol:chdilatonic}
\ph(r) &=&  \fr{1}{6} \log \ls 1 + \fr{Q^2 }{8 m r^2} \rs    , \nn
A(r) &=&  \fr{1}{3} \log \ls 1 + \fr{Q^2 }{8 m r^2} \rs    , \nn
B(r) &=&  - \fr{2}{3} \log \ls 1 + \fr{Q^2}{8 m r^2} \rs    , \nn
f(r)  &=& 1 - \fr{m}{  r^4 \ls 1 + \fr{Q^2}{8 m r^2 } \rs^2 } , \nn
A_t  &=& 2 \k^2 \m - \fr{Q  }{2 r^2 \ls 1 + \fr{Q^2}{8 m  r^2} \rs } ,
\eea
where $m$, $\m$, and $Q$ indicate the charged black brane's mass, chemical potential, and charge density, respectively.
Using the regularity of $A_t$ at the event horizon, the charge density and black brane mass can be rewritten in terms of temperature and chemical potential
\bea   \la{rel:mqtm}
Q &=&  4 \pi^2  \k^2  T_H^2 \m + 8 \k^6 \m^3   , \nn
\sqrt{m} &=& \pi^2 T_H^2 + 2 \k^4 \m^2 .
\eea

Let us first summarize thermodynamic properties of the charged dilatonic black brane. For $m = \fr{Q^{4/3}}{4}$, the above charged dilatonic black brane has an extremal limit in which the horizon resides at $r_h = 0$. Absence of a conical singularity at the event horizon yields the Hawking temperature 
\be
T_H = \fr{r_h}{ \pi } .
\ee
This result shows that the  Bekenstein-Hawking entropy automatically vanishes at zero temperature. It is worth noting that, since the curvature scalar in the extremal limit diverges at the center, the charged dilatonic black brane geometry is incomplete at zero temperature. Nevertheless, the thermodynamic quantities are well-defined even at zero temperature because the divergences of the Einstein-Hilbert and scalar kinetic terms are exactly canceled. In the grand canonical ensemble described by
the following grand potential
\be
\O = - \fr{\pi^4 V_3}{2 \k^2} T_H^4 - 2 \pi^2 \k^2 V_3 T_H^2 \m^2 - \fr{10}{3} \k^6 V_3 \m^4 ,
\ee
other thermodynamic quantities satisfying the first law of thermodynamics are given by \cite{Park:2014gja} 
\bea
E &=& \fr{3 \pi^4 V_3}{2 \k^2} T_H^4 + 6 \pi^2 \k^2 V_3 T_H^2 \m^2 + \fr{14}{3} \k^6 V_3 \m^4 ,
\la{res:internalen}\\
P &=& \fr{ \pi^4 }{2 \k^2} T_H^4 + 2 \pi^2 \k^2 T_H^2 \m^2 + \fr{10}{3} \k^6  \m^4 , \\
S_{BH} &=& \fr{2 \pi^4 V_3}{\k^2} T_H^3 + 4 \pi^2 \k^2 V_3 T_H  \m^2 , \\
\fr{N}{V_3} &=& 4 \pi^2 \k^2 T_H^2 \m + \fr{40}{3} \k^6 \m^3 \la{res:nuberden} ,
\eea
where $E$, $P$, $S_{BH}$ and $N/V_3$ indicate the energy, pressure, entropy and charge density respectively. 

In the AdS/CFT context, they can be reinterpreted as those of the dual field theory. At low temperature, the thermal entropy is linearly proportional to temperature. Furthermore, the extremal limit supports normal modes of massless charged fermions so that the geometry we consider is regarded as the dual of a Fermi liquid \cite{Gubser:2009qt}. The internal energy and pressure correspond to the energy-momentum tensor of the dual theory. Taking the trace, we finally arrive at
\be
{T^{\m}}_{\m} = E - 3 P V_3 = \fr{16}{3} \k^6 V_3 \m^4 .
\ee
This result shows that the trace of the stress tensor does not vanish even though the asymptotic geometry is given by the AdS space. This is because the dual matter we consider is non-conformal. The bulk gauge field is dual to a fermionic number operator \cite{Lee:2009bya,Lee:2015rxa} and the scalar field plays the role of a dilaton because it controls the physical gauge coupling \cite{Gubser:2009qt}. This fact implies that dual matter interacts with gauge bosons nontrivially and that its coupling constant nontrivially runs along the RG flow.

\section{Thermodynamics-like law of the entanglement entropy}

\subsection{entanglement entropy bound in a strip region}
Let us first take into account an entanglement entropy in a strip-shaped region, Assuming that the total system resides in a regularized volume
\be
\fr{L}{2} \le x_1, x_2,  x_3 \le \fr{L}{2} ,
\ee
and that it is divided into two subsystems, $A$ and $\bar{A}$,  then the entanglement entropy of the subsystem $A$ is given by the trace of a reduced density matrix. Parameterizing $A$ as
\be
\fr{l}{2} \le x_1 \le \fr{l}{2} \quad  {\rm and} \ \fr{L}{2} \le  x_2,  x_3 \le \fr{L}{2} ,
\ee
the entanglement entropy can be alternatively evaluated by using the AdS/CFT correspondence. The entanglement entropy is proportional to the area of the minimal surface whose end coincides with the entangling surface we chose. Using the previous metric solution, the minimal surface is given by
\be
A = L^2 \int dx_1 \ \fr{e^{2 A}  \sqrt{e^{2 A} f+ e^{ 2 B} z'^2}}{z^3 \sqrt{f}}  ,
\ee
where $z=1/r$ and the prime indicates a derivative with respect to $x_1$. Since the above action does not depend on $x_1$ explicitly, there exists a conserved quantity 
\be
H = - \fr{e^{4 A} \sqrt{f}}{z^3  \sqrt{e^{2 A} f + e^{ 2 B} z'^2}  } .
\ee
Moreover, the minimal surface is invariant under $x_1 \to - x_1$ so that it should have an extremum point at $x_1 = 0$ which we call a turning point denoted by $z_*$. At the turning point, the conserved quantity reduces to
\be
H = - \fr{ e^{3 A_*}}{ z_*^3} ,
\ee
where the subscript $*$ means the value at the turning point. Comparing these two conserved quantities in the UV region, we can expand $l$ up to $z_*^5$ order
\bea
l = \fr{\G \ls \fr{2}{3}\rs \G \ls \fr{5}{6}\rs}{\sqrt{\pi}} z_*
- \fr{4}{3} \k^4 \m^2 z_*^3
+ \fr{4 \G \ls \fr{1}{3}\rs \G \ls \fr{1}{6}\rs}{15 \sqrt{\pi}}  \k^8 \m^4 z_*^5 
+ \fr{\G \ls \fr{1}{3}\rs \G \ls \fr{1}{6}\rs}{30 \sqrt{\pi}}  \ls \pi^2 T_H^2 + 2 \k^4 \m^2 \rs^2  z_*^5 + \cdots .
\eea
Here, the leading term comes from the pure AdS geometry and the second and third terms correspond to the first and second order corrections originated from $e^{2A}$ and $e^{2B}$. The last term is another second order correction originated from the black brane metric factor. Rewriting the turning point in terms of $l$, we arrive at
\bea
z_* &=&  \fr{\sqrt{\pi}}{\G \ls \fr{2}{3}\rs \G \ls \fr{5}{6} \rs} l + \fr{4 \pi^2 }{3 \ \G \ls \fr{2}{3}\rs^4 \G \ls \fr{5}{6}\rs^4} \k^4  \m^2 l^3   + \fr{4 \pi^{5/2}  \lc 60 \pi - \G \ls \fr{1}{3}\rs \G \ls \fr{1}{6}\rs
\G \ls \fr{2}{3}\rs \G \ls \fr{5}{6}\rs \rc}{45 \  \G \ls \fr{2}{3}\rs^7 \G \ls \fr{5}{6}\rs^7}   \k^8 \m^4 l^5 \nn
&& - \fr{ \pi^{5/2}  \ \G \ls \fr{1}{3}\rs \G \ls \fr{1}{6}\rs}{30 \ \G \ls \fr{2}{3}\rs^6 \G \ls \fr{5}{6}\rs^6}  \ls \pi^2 T_H^2 + 2 \k^4 \m^2 \rs^2  l^5 + \cdots .
\eea

Using these results, the entanglement entropy defined by $ S\equiv \fr{2 \pi A}{\k^2}$ becomes in terms of $l$
\bea
&& S \ls T_H, \m \rs = \fr{2 \pi}{\k^2} \fr{L^2}{\e^2}  - \fr{\G \ls \fr{2}{3}\rs^3 \G \ls \fr{5}{6}\rs^3 }{\sqrt{\pi} \k^2 } \fr{L^2 }{l^2} \nn
&& \qquad  +  \fr{8 \pi }{3 \k^2} L^2 \kappa^4 \mu ^2
+ \fr{ 16 \pi^{5/2} \ls 15 -  2 \sqrt{3} \pi \rs}{45 \k^2 \ \G \ls \fr{2}{3}\rs^3 \G \ls \fr{5}{6}\rs^3} L^2 \k^8 \m^4  l^2  + \fr{ 2 \pi^{7/2}  \ls \pi^2 T_H^2 + 2 \k^4 \m^2 \rs^2 L^2  l^2}{5  \sqrt{3} \k^2 \ \G \ls \fr{2}{3}\rs ^3 \G \ls \fr{5}{6}\rs^3}   + \cdots ,
\eea
where the ellipsis means higher order corrections. Above the first line is the entanglement entropy of the pure AdS space, while the rest denote contributions from the metric components and the black brane factor.  At zero temperature with a nonzero chemical potential, the ground state has the following entanglement entropy
\bea
S \ls 0, \m \rs &=& \fr{2 \pi}{\k^2} \fr{L^2}{\e^2}  - \fr{\G \ls \fr{2}{3}\rs^3 \G \ls \fr{5}{6}\rs^3 }{\sqrt{\pi} \k^2 } \fr{L^2 }{l^2} +  \fr{8 \pi }{3 \k^2} L^2 \kappa^4 \mu ^2  + \fr{ 8 \pi^{5/2} \ls 30 -  \sqrt{3} \pi \rs}{45  \ \G \ls \fr{2}{3}\rs^3 \G \ls \fr{5}{6}\rs^3} L^2 \k^6 \m^4  l^2    + \cdots ,
\eea
which differs from that of the vacuum because the ground state is already occupied by matter. When the ground state is excited without the change of $\m$, the increased entanglement entropy is given by
\bea
\lp \D S \right|_{\m} &\equiv& S \ls T_H, \m \rs  - S \ls 0, \m \rs
= \fr{ 2 \ \pi^{11/2}  L^2  l^2  T_H^2 }{5  \sqrt{3} \k^2 \ \G \ls \fr{2}{3}\rs ^3 \G \ls \fr{5}{6}\rs^3} \left(\pi^2 T_H^2+4  \kappa ^4 \mu ^2 \right)   + \cdots .
\eea
According to the entanglement entropy bound, the increase of the entanglement entropy should be bounded by the increased energy. Especially when only low order corrections are taken into account, the entanglement entropy bound is saturated. In order to check this point, let us calculate the increased energy when the ground state is excited. At a given chemical potential, the energy used to excite the ground state is evaluated from \eq{res:internalen}, which can be reinterpreted as the energy density of excited states in a small subsystem \cite{Bhattacharya:2012mi},
\be		
\lp \D E \right|_{\m} \equiv E \ls T_H, \m \rs  - E \ls 0, \m \rs = \frac{3 \pi ^2 l L^2  T_H^2}{2 \kappa ^2} \left(\pi ^2 T_H^2 + 4 \kappa ^4 \mu ^2 \right) ,
\ee
where $l L^2$ corresponds to the volume of the subsystem. Note that this increased energy is exact because there are no more higher order corrections.

If we consider only the $l^2$ order correction in the above entanglement entropy, the increased energy and entanglement entropy satisfy the following relation
\be    \la{re:Bebound}
\lp \D S \right|_{\m} =   \fr{ 4 \pi^{7/2} }{15  \sqrt{3} \ \G \ls \fr{2}{3}\rs ^3 \G \ls \fr{5}{6}\rs^3} \ l  \lp \D E \right|_{\m} ,
\ee
which is the form when the Bekenstein bound is saturated. Unfortunately, since it has not been known how to calculate the modular Hamiltonian in a strip-shaped region we cannot directly compare this result with the entanglement entropy bound. In spite of this fact, we can guess from \eq{ass:formofmodHam} that the modular Hamiltonian in the strip region should be given by the right hand side of \eq{re:Bebound}. Intriguingly, the above result can be reinterpreted as the thermodynamics-like law, $\lp \D E  \right|_{\m} = T_E \lp \D S\right|_{\m} $. To do so, we should define an entanglement temperature inversely proportional to the strip width
\be
T_E =  \fr{15  \sqrt{3} \ \G \ls \fr{2}{3}\rs ^3 \G \ls \fr{5}{6}\rs^3}{ 4 \pi^{7/2} } \fr{1}{l} .
\ee
The thermodynamic interpretation and definition of the entanglement temperature are meaningful only when the Bekenstein bound is saturated. If we consider higher order corrections, the Bekenstein bound implies that the increased entanglement entropy should be smaller than the increased energy
\be
\lp \D S  \right|_{\m}  < \fr{ 4 \pi^{7/2} }{15  \sqrt{3} \ \G \ls \fr{2}{3}\rs ^3 \G \ls \fr{5}{6}\rs^3} \ l  \lp \D E \right|_{\m}  .
\ee

\subsection{Entanglement entropy bound in a ball-shaped region}

In general, the modular Hamiltonian is a complicated object which cannot be expressed as an integral of local operator
except several simple cases. One of the exceptions appears when one considers a spherical entangling surface. If the quantum state is excited without the change of the chemical potential,
the modular Hamiltonian is associated with the stress tensor \cite{Casini:2011kv,Faulkner:2013ica}
\bea			\la{rel:modandenergy}
\lp K \right|_{\m}  &=& 2 \pi \O_2 \int_{\r \le l} d\r \ \r^2 \ \fr{l^2 - \r^2}{2 l}  \lp T_{00}  \right|_{\m} 
\eea
where $\lp T_{00} \right|_{\m} $ indicates the energy density at a given $\m$ and $\O_2$ is the solid angle of the spherical entangling surface. Since $T_{00}$ is uniform, the modular Hamiltonian can be rewritten as
\be
 \lp K \right|_{\m}   = \fr{2 \pi }{5} l  \lp E \right|_{\m}  ,
\ee
where the energy contained in the ball-shaped region is given by $\lp E \right|_{\m}  = \O_2 \int_{\r \le l} d\r \ \r^2 \lp T_{00} \right|_{\m}  $. 
This relation shows how the modular Hamiltonian is related to the energy over the interior of the sphere. This is the form expected in \eq{ass:formofmodHam} and shows that the entanglement entropy bound is equivalent to the Bekenstein bound except that the former is also working in a quantum system. Substituting the energy obtained from the black brane thermodynamics, the explicit modular Hamiltonian reads
\be  		\la{res:modularham}
\lp  K \right|_{\m}  = \fr{\pi l^4 \O_2}{5 \k^2}   \left(\pi ^2 T_H^2+2 \kappa ^4 \mu ^2\right)^2
- \frac{8 \pi l^4 \Omega _2}{45 \kappa ^2 }      \kappa^8  \mu ^4 .
\ee
When $\m$ is fixed, the increased modular Hamiltonian becomes
\be			\la{res:incrementenergy}
\lp \D K \right|_{\m}   \equiv K \ls T_H, \m \rs -   K \ls 0, \m \rs= \fr{\pi^5 l^4 \O_2}{5 \k^2}  T_H^4
+ \fr{4 \pi^3 \kappa ^4  l^4 \O_2}{5 \k^2}  \mu ^2  T_H^2 .
\ee
When $\m=0$, it reduces to that of the Schwarzschild AdS black brane. As mentioned before, the non-negativity of the relative entropy implies that the increase of the entanglement entropy is bounded by the increased modular Hamiltonian, $\lp \D S \right|_{\m}  \le \D \lp K \right|_{\m}  $.

Now, let us consider the entanglement entropy contained in a ball-shaped region. Parameterizing a disk with a radius $l$ as
\be
0 \le \r \le l .
\ee
and considering $z$ as a function of $\r$, the action for the minimal surface is reduced to
\be         \la{act:diskminsurf}
A =  \O_2 \int_0^{l-l_*} d \r \ \fr{e^{2 A} \r^2 }{z^3 } \sqrt{e^{2 A} + \fr{e^{2 B}}{f} z'^2} ,
\ee
where $l_*$ in the upper limit is introduced to denote a UV cutoff and the prime means a derivative with respect to $\r$. The equation of motion for $z$ reads
\bea            \la{eq:eominsphere}
0 &=& \rho  z f z'' +2 z  z'^3 e^{2 B -2 A}-4 \rho  z  f z'^2 A' +\rho  z  f  z'^2 B' -\frac{1}{2} \rho  z z'^2 f'\nn
&& -3 \rho  z f ^2 e^{2 A -2 B } A'
+3 \rho  f ^2 e^{2 A -2 B } +3 \rho  f z'^2+2 z  f  z'  .
\eea
Since $\m l$ and $T_H l$ have small values in a UV region, one can expand $z$ as follows
\be
z (\r) = z_0 (\r) + \k^4 \m^2 l^2 z_1 (\r)  +
\k^8 \m^4 l^4 z_2 (\r)  + T_H^4 l^4 z_3 (\r)  +  \k^4 \m^2 T_H^2 l^4  z_4  (\r)  + \cdots ,
\ee
where the ellipsis indicates higher order corrections. Related to $T_H$, note that the lowest corrections appear as the forms, $T_H^4 l^4$ and $T_H^2 \m^2 l^4$, because there is no $T_H^2 l^2$ term in \eq{eq:eominsphere}. This is the reason why our ansatz does not include a $T_H^2 l^2$ term. However, since $e^{2A}$ and $e^{2B}$ contain terms proportional to $\m^2 l^2$, the ansatz we have taken should have a $\m^2 l^2$ term in order to satisfy the equation of motion.

At leading order, the action is exactly reduced to that obtained from a pure AdS space and its solution has already been known as \cite{Ryu:2006bv,Ryu:2006ef}
\be
z_0 (\r)  = \sqrt{l^2 - \r^2} .
\ee
Around this known solution, the first correction caused by the deformation is governed by $z_1$. At $l^2$ order, $z_1$ is given by
\be
z_1 = - \fr{2 \r^2 (2 l^2 - \r^2)}{3 l^2 \sqrt{l^2 - \r^2}} +
\fr{(l - \r)^2 c_1}{\r \sqrt{l^2 - \r^2}}
+ \fr{c_2}{\sqrt{l^2 - \r^2}} .
\ee
Since the subsystem we consider is located at $z=0$, all higher order functions should vanish at $\r=l$. This constraint fixes $c_2$ to be $c_2 = \fr{2 l^2}{3}$. In addition, the smoothness of the minimal surface at the turning point, $z_1'=0$ at $\r=0$, determines the remaining integral constant to be $c_1=0$. Substituting this solution back into the action in \eq{act:diskminsurf}, one can obtain two different corrections caused by the metric and minimal surface deformations at $l^2$ order.

At $l^4$ order, $z_2$, $z_3$ and $z_4$ satisfying equations of motion are given by
\bea
z_2 &=& \fr{2  (5 l^4 \r^2 + 2 l^2 \r^4 - 3 \r^6)}{ 45 l^4 \sqrt{l^2 - \r^2}}
+ \fr{(l - \r)^2 c_3}{\r \sqrt{l^2 - \r^2} }+  \fr{c_4}{\sqrt{l^2 - \r^2} }  , \nn
z_3 &=& \fr{\pi^4 (5 l^4 \r^2 - 4 l^2 \r^4 + \r^6)}{ 10 l^4 \sqrt{l^2 - \r^2}}
+ \fr{(l - \r)^2 c_5}{\r \sqrt{l^2 - \r^2} }+  \fr{c_6}{\sqrt{l^2 - \r^2} }  , \nn
z_4 &=& \fr{2 \pi^2 (5 l^4 \r^2 - 4 l^2 \r^4 + \r^6)}{ 5 l^4 \sqrt{l^2 - \r^2}}
+ \fr{(l - \r)^2 c_7}{\r \sqrt{l^2 - \r^2} }+  \fr{c_8}{\sqrt{l^2 - \r^2} } .
\eea
Imposing again that all higher order functions should be zero at $\r=l$, half of unknown integral constants are determined to be
\be
c_2 = \fr{2 l^2}{3} \quad , \quad c_4 = - \fr{8 l^2}{45} \quad , \quad c_6 = - \fr{\pi^4 l^2}{5} \quad {\rm and} \quad
c_8 = - \fr{4 \pi^2 l^2}{45} .
\ee
In addition, the smoothness of the minimal surface at the turning point, $z_1' = z_2' =z_3' =z_4' =0$ at $\r=0$, yields $c_1=c_3=c_5=c_7=0$. Substituting the above solutions into the action, the minimal area up to $l^4$ order gives rise to
\be
A = \O_2 \int_0^{l-l_*} d \r  \lb  \frac{l \rho ^2  }{\left(l^2-\rho ^2\right)^2}-\frac{8}{15} \kappa ^8 l
\mu ^4 \rho ^2   + \frac{3}{5} l \rho ^2   \left(\pi ^2 T_H^2+2 \kappa ^4 \mu ^2\right)^2  \rb .
\ee
Here $l_*$ is associated with the UV cutoff of the $z$ coordinate denoted by $\e$. The perturbative solution we found determines their relation up to higher order corrections:
\be
l_* =  \fr{\e^2}{2 l} \lb 1 +
\lc \frac{1}{4 l^2} -\frac{4 \kappa ^4 }{3} \mu ^2 -\frac{8  l^2}{45} \kappa ^8 \mu ^4
+ \frac{l^2}{5}  \left(\pi ^2 T_H^2+2 \kappa ^4 \mu ^2\right)^2 \rc \e^2 + \cdots \rb .
\ee
Using this relation, the entanglement entropy finally becomes
\bea		\la{res:entanglementdisk}
S \ls T_H,\m \rs &=& \frac{\pi    l^2 \Omega _2}{\kappa ^2 \epsilon ^2}
+\frac{\pi  \Omega _2 }{\kappa ^2} \log \ls \fr{\epsilon}{l} \rs
 -\frac{\pi  \Omega _2}{2 \kappa ^2} \ls 1 + 2 \log 2 \rs  \nn
&&  +  \frac{4 \pi  l^2  \Omega _2}{3 \kappa ^2 }   \kappa ^4 \mu ^2
-\frac{8 \pi l^4 \Omega _2}{45 \kappa ^2 }      \kappa^8  \mu ^4
+   \frac{\pi   l^4 \Omega _2 }{5   \kappa ^2} \left(\pi ^2 T_H^2+2 \kappa ^4 \mu ^2\right)^2 .
\eea
This is the entanglement entropy of the excited state with the chemical potential. 

At a given chemical potential, the increased entanglement entropy up to $l^4$ order is given by
\be
\lp \D S   \right|_{\m}  \equiv S \ls T_H,\m \rs -  S \ls 0,\m \rs = \fr{\pi^5 l^4 \O_2}{5 \k^2}  T_H^4
+ \fr{4 \pi^3 \kappa ^4  l^4 \O_2}{5 \k^2}  \mu ^2  T_H^2 ,
\ee
which is the exact same as the increased modular Hamiltonian in \eq{res:incrementenergy}. When higher order corrections are ignored, the almost saturated entanglement entropy bound leads to the thermodynamics-like law
\be		\la{res:eebforex}
\lp \D  K \right|_{\m}  = \lp \D S  \right|_{\m}  =\fr{1}{T_E}    \lp  \D E \right|_{\m} ,
\ee
with 
\be
T_E =  \fr{5}{2 \pi l}  .
\ee  
As mentioned before, the entanglement temperature shows a universal feature proportional to the inverse of the system size.  
In order to understand this result, let us first consider black hole thermodynamics. In general, a charged black hole has an additional conserved charge and its thermodynamics 
\be			\la{res:gBB}
d E = T_H dS_{BH} + \m d N ,
\ee
can be identified with that of the dual field theory in the AdS/CFT contexts. 
 When an additional neutral particle is absorbed into the charged black hole, the energy and entropy usually increase. However, the charge does not because a neutral particle has no charge. This means $dN=0$, so the corresponding thermodynamic law is reduced to
\be			\la{res:thermolaw1}
d E = T_H dS_{BH}  .
\ee
Comparing it with the above entanglement entropy bound in \eq{res:eebforex}, it is similar to the entanglement entropy bound. Since the entanglement entropy bound is regarded as the quantum generalization of the Bekenstein bound, the entanglement entropy bound, \eq{res:eebforex}, reduces to the Bekenstein bound, \eq{res:thermolaw1}, in the IR limit where the entanglement entropy yields the Bekenstein-Hawking entropy. On the other hand, finite thermal fluctuations can be ignored in the UV limit, so that $ \lp  \D E \right|_{\m}$ can be regarded as the quantum excitation energy. This quantum excitation energy increases the entanglement entropy and their ratio plays the role of temperature according to the analogy to the first law of thermodynamics, which was called the entanglement temperature to distinguish it from the normal temperature.

Now, let us consider a global quench corresponding to a sudden chemical potential change at a given $T_H$. To do so, it is more convenient to consider $N$ as a fundamental variable instead of the chemical potential. This is associated with the Legendre transformation and the change of the chemical potential is due to the change of the particle number. 
From \eq{res:nuberden}, the chemical potential can be written as a function of $N$
\be
\m=\frac{2^{1/3} \left(  \sqrt{160 \pi ^6 V_3^4 T_H^6+ 225 N^2}+15 N \right){}^{2/3}-4 \ 5^{1/3}  \pi ^2 V_3^{4/3} T_H^2}{2\ 10^{2/3}
   \kappa ^2 V_3^{2/3}   \left(  \sqrt{160 \pi ^6 V_3^4 T_H^6+ 225 N^2}+15 N  \right)^{1/3}} \, ,
\ee
where $V_3$ indicates the volume of the ball, $V_3 = \fr{ l^3}{3} \O_2$. 
When the particle number is slightly changed ($\D N \ll N$), the change of the chemical potential is given by 
\be
\lp \D \m \right|_{T_H}   = \lp \fr{\pa \m}{\pa N} \right|_{T_H}  \D N ,
\ee
with
\bea
\lp \fr{\pa \m}{\pa N} \right|_{T_H} &=& \fr{ \ 5^{1/3} \left(\sqrt{32 \pi ^6 V_3^4 T_H^6+ 45 N^2}+ 3 \sqrt{5} N\right) }{2 \kappa ^2  V_3^{2/3} \sqrt{32 \pi ^6 V_3^4 T_H^6+ 45 N^2}} \nn
&& \times \  \frac{\left(\sqrt{160 \pi ^6 V_3^4   T_H^6+ 225 N^2}+ 30 N\right){}^{2/3}+2 \ 10^{1/3} \pi ^2 V_3^{4/3} T_H^2 }{ \left( \sqrt{160 \pi ^6 V_3^4 T_H^6+ 225 N^2 }+ 15 N \right){}^{4/3} }  .
\eea
In addition, this global quench also leads to the change of the energy at a given $T_H$
\be  		\la{res:exciteMH}
\fr{\lp \D E \right|_{T_H}}{T_E}   = \fr{8 \pi^3   \kappa ^4  l^4 \O_2}{5 \k^2}   \pi ^2 \m T_H^2  \lp \D \m \right|_{T_H} 
+\fr{ 112 \pi l^4 \O_2}{45 \k^2}  \kappa ^8 \mu ^3  \lp \D \m \right|_{T_H}  ,
\ee
and the increased entanglement entropy up to $l^4$ order reads 
\be		\la{res:exciteEE}
\lp \D S \right|_{T_H}=   \frac{8 \pi  l^2  \Omega _2}{3 \kappa ^2 }   \kappa ^4 \mu  \lp \D \m \right|_{T_H} 
+ \fr{4 \pi^3   \kappa ^4  l^4 \O_2}{5 \k^2}    \pi ^2 \mu T_H^2 \lp \D \m \right|_{T_H} 
+\fr{ 112 \pi l^4 \O_2}{45 \k^2}  \kappa ^8 \mu ^3 \lp \D \m \right|_{T_H} .
\ee

Comparing these two results, the thermodynamics-like law in \eq{res:eebforex} is violated under a global quench. To understand why this happens, let us first consider the thermodynamic law of a charged black hole. When a charged particle instead of a neutral one is absorbed, the energy and entropy as well as the charge of the black hole are changed. These
quantities satisfy the generalized first law of thermodynamics in \eq{res:gBB}. This relation implies that adding more particles modifies the thermodynamic law. Relying on the charge of the absorbed particle, the chemical potential can have a positive or negative value. 
Similarly, we also expect that the entanglement entropy bound is modified under a global quench. Suppose that $K_0$ is the modular Hamiltonian of the undeformed theory. Then, the reduced density matrix is given by
\be
\r_0 = \fr{e^{- K_0}}{\Tr e^{- K_0}} .
\ee
In a ball-shaped region, the modular Hamiltonian is related to the energy,  $ K_0 = \fr{ E}{T_E}$ from \eq{rel:modandenergy}.
Now, let us deform this theory by a relevant number operator, $N$,
\be
K = K_0 - \fr{\m_E}{T_E}  N , 
\ee
where the entanglement chemical potential, $\m_E$, accounts for how a global quench modifies the modular Hamiltonian and entanglement entropy \cite{Park:2015afa}. The reduced density matrix of the deformed theory becomes
\be
\r = \fr{e^{- K}}{\Tr e^{- K}}
\ee
and the non-negativity of the relative entropy gives rise to a generalized entanglement entropy bound
\be
\D K =  \fr{ \D E}{T_E} - \fr{\m_E}{T_E}  \D N \ge \D S .
\ee
Note that the previous result in \eq{res:eebforex} is a special case with $\D N =0$.

When the generalized entanglement entropy bound is saturated, its form is the same as the generalized first law of thermodynamics in \eq{res:gBB}. From  \eq{res:exciteMH} and \eq{res:exciteEE}, the entanglement chemical potential is given by
\be
\m_E= - \frac{5^{1/6} l \Omega _2 \sqrt{32 \pi ^6 V_3^4 T_H^6+ 45 N^2 } \left[\left( 
   \sqrt{160 \pi ^6 V_3^4 T_H^6+ 225 N^2 }+15 N\right){}^{4/3} - 8 \ 2^{1/3} 5^{2/3} \pi ^4 V_3^{8/3} T_H^4 \right]}
   {3 \kappa ^2 V_3^{1/3}  \left(32 \pi ^6 V_3^4 T_H^6+45 N^2\right) \left(2    \sqrt{160 \pi ^6 V_3^4 T_H^6+ 225 N^2 }+30 N\right){}^{2/3}} ,
\ee
where $\m_E \le 0$. Unlike the entanglement temperature, the entanglement chemical potential usually has a nontrivial 
size dependence \cite{Park:2015afa}.
Assuming that one can substitute more particles without changing the energy, then the entanglement entropy change from the generalized entanglement entropy bound becomes
\be
\D S =  - \fr{\m_E}{ T_E  }   \D N .
\ee
Since $\m_E$ is negative, adding more particles increases the entanglement entropy as expected. As a consequence, the generalized entanglement entropy bound is still satisfied under a global quench.


\section{Discussion}

When an object is absorbed into a black hole, the Bekenstein bound has been proposed to explain the increase of the thermal entropy. In this paper, we have investigated the generalized entanglement entropy bound for a holographic fermion system with a Fermi surface. Intriguingly,  the entanglement entropy bound is originated from the non-negativity of the relative entropy and can be applied to a quantum system unlike the Bekenstein bound.  Rewriting the entanglement entropy bound in terms of the system energy instead of the modular Hamiltonian, it is equivalent to the Bekenstein bound except that it is working even in a quantum system. Recently, it has been shown that the entanglement temperature satisfying the thermodynamics-like law has a universal feature inversely proportional to the system size \cite{Bhattacharya:2012mi}. We showed that the almost saturated entanglement entropy bound can account for the universality of the entanglement temperature. 

We have also studied how the entanglement entropy bound is modified under a global quench. In general, a global quench changes the modular Hamiltonian and entanglement entropy which lead to the generalized entanglement entropy bound. When the generalized entanglement entropy bound is saturated, we showed that it also satisfies the generalized thermodynamic law. Unlike the entanglement temperature, the entanglement chemical potential nontrivially depends on the system size.

\vspace{1cm}

{\bf Acknowledgement}

C. Park was supported by Basic Science Research Program through the National Research Foundation of Korea funded by the Ministry of Education (NRF-2013R1A1A2A10057490) and also by the Korea Ministry of Education, Science and Technology, Gyeongsangbuk-Do and Pohang City.


\end{document}